\documentclass[12pt]{article}

\usepackage[USenglish]{babel} 
\usepackage[T1]{fontenc}
\usepackage[ansinew]{inputenc}
\usepackage{verbatim}
\usepackage[sort,move]{cite}

\usepackage{lmodern} 
\usepackage{color}

\usepackage{graphicx} 

\usepackage{amsmath}
\usepackage{amsthm}
\usepackage{amsfonts}

\usepackage{hyperref}

\def\a{\alpha}
\def\b{\beta}
\def\c{\gamma}
\def\d{\delta}
\def\e{\epsilon}           
\def\g{\gamma}

\def\m{\mu}
\def\n{\nu}
  
\def\p{\pi}                
\def\r{\rho}                                     
\def\s{\sigma}                                   
\def\t{\tau}

\def\D{\Delta}

\def\S{\Sigma}

\def\del{\partial}              
\let\a=\alpha \let\b=\beta \let\g=\gamma \let\d=\delta \let\e=\epsilon
    
 \let\m=\mu \let\n=\nu  \let\r=\rho
\let\s=\sigma \let\t=\tau   \let\c=\chi 
  \let\D=\Delta  
    
\let\ep=\epsilon

\def\nn{\nonumber} \def\bd{\begin{document}} \def\ed{\end{document}}
\def\ds{\documentstyle} \let\fr=\frac \let\bl=\bigl \let\br=\bigr
\let\Br=\Bigr \let\Bl=\Bigl
\let\bm=\bibitem
\let\na=\nabla
\let\pa=\partial \let\ov=\overline
\newcommand{\be}{\begin{equation}}
\newcommand{\ee}{\end{equation}}
\def\ba{\begin{array}}
\def\ea{\end{array}}
\def\ft#1#2{{\textstyle{{\scriptstyle #1}\over {\scriptstyle #2}}}}
\def\fft#1#2{{#1 \over #2}}
\def\del{\partial}
\def\sst#1{{\scriptscriptstyle #1}}
 \def\oneone{\rlap 1\mkern4mu{\rm l}}
\def\ie{{\it i.e.\ }}
\def\via{{\it via}}
\def\semi{{\ltimes}}
\def\str{{\rm str}}
\def\tr{{\rm tr}}
\def\Dm{{{D_{\sst{max}}}}}
\def\vac{ \left | 0 \right \rangle }
\def\kvac{ \left | k \right \rangle }

\def\H{\mathcal{H}}
\def\red{\color{red}}
\def\sp{\; \; \;}

\def\bol{ \left | B (p^+) \right \rangle}
\def\bo1{ \left | B^0 (p^+) \right \rangle}

\def\bolt{ \left | B (p^+) \right \rangle_{\t}}

\def\boxl{ \left | B (x^-) \right \rangle}
\newcommand{\bea}{\begin{eqnarray}}
\newcommand{\eea}{\end{eqnarray}}

\def\<{ \langle }
\def\>{ \rangle }

\def\S{\Sigma}
\renewcommand{\floatpagefraction}{0.6}
\renewcommand{\textfraction}{0.2}

\newcommand\ca{\mathcal{A}}
\newcommand\vp{\varphi}
\newcommand\beal{\begin{align}}
\newcommand\bbone{\ensuremath{\mathbbm{1}}}

\newcommand{\eq}[1]{\begin{equation}#1\end{equation}}
\newcommand{\spl}[1]{\begin{split}#1\end{split}}
\newcommand{\al}[1]{\begin{align}#1\end{align}}
\newcommand{\subeq}[1]{\begin{subequations}#1\end{subequations}}

\newcommand{\arXividhepth}[1]{\href{http://arxiv.org/abs/#1}arXiv:{\tt #1} [hep-th]}
\newcommand{\arXividother}[2]{\href{http://arxiv.org/abs/#1}arXiv:{\tt #1} [#2]}

\newcommand{\bg}[1]{\hat{#1}}
\newcommand{\wj}{\widetilde{J}}
\newcommand{\reo}{\mathrm{Re}~\!\omega}
\newcommand{\imo}{\mathrm{Im}~\!\omega}
\newcommand{\ads}{AdS_4}
\newcommand{\mcal}{\mathcal{M}}
\newcommand{\ccal}{\mathcal{C}}
\newcommand{\ncal}{\mathcal{N}}
\newcommand{\boxedeq}[1]{
\begin{equation}
\fbox{
\rule[0.7cm]{0pt}{0pt}
$#1$
\rule[-0.45cm]{0pt}{0pt}
}
\end{equation}
}

\def\d{\text{d}}

\def\slashchar#1{\setbox0=\hbox{$#1$}           
\dimen0=\wd0                                 
\setbox1=\hbox{/} \dimen1=\wd1               
\ifdim\dimen0>\dimen1                        
\rlap{\hbox to \dimen0{\hfil/\hfil}}      
#1                                        
\else                                        
\rlap{\hbox to \dimen1{\hfil$#1$\hfil}}   
/                                         
\fi}

\def\Re           {{\rm Re\hskip0.1em}}
\def\Im           {{\rm Im\hskip0.1em}}

\newcommand{\E}{\text{\tiny E}}

\begin{document}

\begin{titlepage}

\begin{center}


\vskip 2cm
{\Large \bf Lifshitz from AdS at finite temperature and top down models}


\vskip 1.25 cm {\bf Yegor Korovin$^1$, Kostas Skenderis$^{1,2,3}$ and Marika Taylor$^{2,3}$}
\\ {\vskip 0.5cm \it\small
KdV Institute for Mathematics$^1$,
Institute for Theoretical Physics$^2$, \\
Science Park 904, 1090 GL Amsterdam, the Netherlands. \\
\vskip 0.2cm
Mathematical Sciences and STAG research centre$^3$,  University of Southampton, \\
Highfield, Southampton, SO17 1BJ, UK.}

{\vskip 0.2cm \small {\it E-mail: J.Korovins@uva.nl, K.Skenderis@soton.ac.uk and  M.M.Taylor@soton.ac.uk} }

\end{center}

\vskip 1 cm

\begin{abstract}
\baselineskip=16pt
We construct analytically an asymptotically Lifshitz black brane with dynamical exponent $z=1+\e^2$ in an Einstein-Proca model, where $\e$ is a small parameter. In previous work we showed that the holographic dual QFT is a deformation of a CFT by the time component of a vector operator and the parameter $\e$ is
the corresponding
deformation parameter. In the black brane background this operator additionally acquires a vacuum expectation value.
We explain how the QFT Ward identity associated with Lifshitz invariance leads to a conserved mass and compute analytically the thermodynamic quantities showing
that they indeed take the form implied by Lifshitz invariance. In the second part of the paper  we consider top down Lifshitz models with dynamical exponent close to one and show that they can be understood in terms of vector deformations of conformal field theories. However, in all known cases, both the conformal field theory and its Lifshitz deformations have modes that violate the Breitenlohner-Freedman bound.

\end{abstract}

\end{titlepage}

\setcounter{tocdepth}{2}

\tableofcontents
\pagebreak








\section{Introduction}
In recent years there has been considerable work on the use of holographic models to gain insights into strong coupling physics in condensed matter systems (see \cite{Sachdev:2010ch,McGreevy:2009xe,Horowitz:2010gk,Herzog:2009xv,Hartnoll:2009sz} for reviews). Gauge/gravity duality may be an important tool in understanding strongly interacting non-relativistic scale invariant systems and gravity solutions exhibiting
Schr\"{o}dinger \cite{Balasubramanian:2008dm,Son:2008ye} and Lifshitz symmetry \cite{Kachru:2008yh} have been constructed.

While simple models can capture interesting phenomenology, it is important to understand the nature of the corresponding dual non-relativistic theories better, from first principles. In \cite{Guica:2010sw} (see also \cite{Costa:2010cn,vanRees:2012cw}) it was shown that the field theories dual to Schr\"{o}dinger geometries can be understood as specific deformations of relativistic conformal field theories by operators which are exactly marginal from the perspective of the
Schr\"{o}dinger group, but are irrelevant from the perspective of the conformal symmetry group and moreover break the relativistic symmetry.

Recently an analogous interpretation for Lifshitz spacetimes with dynamical exponent $z$ close to one was developed in \cite{Korovin:2013bua}: a specific deformation of a $d$-dimensional CFT by a dimension $d$ vector operator generically leads to a theory with Lifshitz scaling invariance. For both Schr\"{o}dinger and Lifshitz dualities, this perspective not only elucidates the nature of the non-relativistic theories realised holographically but also demonstrates that new classes of theories with non-relativistic symmetries can be obtained as deformations of relativistic conformal field theories. Since such deformations do not need to be realized holographically, these results are interesting for field theory in their own right and moreover could lead to interesting new weakly interacting non-relativistic theories.

In this paper we consider the finite temperature behaviour of Lifshitz theories in this class, i.e. theories with dynamical exponent $z$ close to one which can be viewed as vector deformations of CFTs, and we also show that top down models in string theory with dynamical exponents close to one indeed lie in this universality class.

 From the bulk perspective, the simplest realization of Lifshitz is the bottom up Einstein-Proca model introduced in \cite{Taylor:2008tg}.
Black hole/brane solutions with Lifshitz asymptotics are needed to study the corresponding dual field theories at non-zero temperature. However, only numerical black hole solutions are available for generic values of $z$  \cite{Danielsson:2009gi,Mann:2009yx,Bertoldi:2009vn, Balasubramanian:2009rx, Pang:2009pd, AyonBeato:2009nh, Cai:2009ac, Cheng:2009df, AyonBeato:2010tm, Balasubramanian:2010uk, Amado:2011nd, Dehghani:2011tx, Chemissany:2011mb, Maeda:2011jj, Matulich:2011ct}.

Note that analytic asymptotically Lifshitz black hole solutions are readily available in Einstein-Dilaton-Maxwell (EDM) theories, see the earliest examples in \cite{Taylor:2008tg,Tarrio:2011de}, with the interpretation of the running scalar being discussed in \cite{Gouteraux:2012yr}. More recently there has been considerable interest in solutions of EDM theories exhibiting hyperscaling violation, see for example
\cite{Dong:2012se,Iizuka:2012iv,Bhattacharya:2012zu,Iizuka:2012pn,Gouteraux:2012yr,Gath:2012pg,Bueno:2012vx}. In this paper we will focus on pure Lifshitz solutions although it would certainly be interesting to understand whether EDM solutions can admit analogous dual interpretations in terms of deformations of relativistic theories and indeed whether EDM solutions can be related to Lifshitz solutions through generalized dimensional reduction \cite{Kanitscheider:2009as,Gouteraux:2011ce,Gouteraux:2011qh}. Note that issues and open questions involving the IR behaviour of the Lifshitz theory, see \cite{Harrison:2012vy} and \cite{Horowitz:2011gh}, do not play a role here.

In the first part of this paper we consider Einstein-Proca models and construct black brane solutions with Lifshitz asymptotics for dynamical exponent $z=1 + \e^2$, with $\e$ being a small expansion parameter. Our solutions are constructed analytically, working perturbatively in $\e$.
Applying the holographic dictionary developed in \cite{Korovin:2013bua} we obtain the one-point function of the dual energy-momentum tensor and check the various thermodynamic relations expected for Lifshitz invariant theories \cite{Bertoldi:2009dt,Bertoldi:2011zr,Barclay:2012he}. In particular we show how the Ward identity due to Lifshitz invariance implies the existence of a conserved mass and we show that the entropy scales with temperature as
\be
\label{ST}
S \propto T^{\frac{d-1}{z}}.
\ee
The thermodynamic quantities are obtained analytically and the analytic solutions could be useful in extracting quasi-normal modes, studying correlation functions etc.

While it is a useful bottom up model, the Einstein-Proca model has a disadvantage: string theory embeddings  are known only for specific values of the dynamical exponent $z$, see for example \cite{Donos:2010ax}, none of which are close to one. There are two main classes of string theory embeddings of Lifshitz solutions known. The first is that of $z=2$ Lifshitz which can be obtained
from reducing $z=0$ Schr\"odinger over a circle \cite{Balasubramanian:2010uk,Costa:2010cn,Narayan:2011az}. This system can be embedded in supergravity \cite{Donos:2010tu,Halmagyi:2011xh,Cassani:2011sv,Petrini:2012bh} and the detailed holographic dictionary was obtained in  \cite{Chemissany:2011mb,Chemissany:2012du}, reducing the results obtained in \cite{Papadimitriou:2011qb}. However the reduction circle becomes null at infinity which implies the dual theory should be related to the Discretized Light Cone Quantization (DLCQ) of the deformed CFT corresponding to the $z=0$ Schr\"odinger solution, and thus
this approach suffers from the well-known subtleties associated with DLCQ. These $z=2$ Lifshitz solutions are not in the same universality class as the solutions discussed in this paper.

The second class of top down embeddings of Lifshitz solutions consist of uplifts of solutions to Romans gauged supergravity theories \cite{Gregory:2010gx}. Lifshitz geometries ${\rm Li}_{D}(z)$ in $D = d +1$ bulk dimensions with generic dynamical exponent $z$ can be realized in this way.
The structure of these solutions is as follows: products of  ${\rm Li}_{D}(z)$ with two-dimensional hyperboloids solve the equations of Romans gauged supergravity theories in $(D+2)$ dimensions, for specific choices of the masses and couplings in these theories.

Since there are Lifshitz solutions with $z \sim 1$ in these top down models, it is interesting to explore whether these can also be understood in terms of deformations of conformal field theories. In section \ref{topdown} we show that these solutions are indeed in the same universality class as the Einstein-Proca model: i.e. to leading order in the parameter $\e$ the dual field theory is a deformation of a CFT by a vector operator. However, unlike the Einstein-Proca model, other CFT operators (which preserve Lifshitz symmetry) are induced at higher orders in $\e$. The bulk theories in this case therefore realize one of the field theory scenarios discussed in section 5 of \cite{Korovin:2013bua}.

Unfortunately not just the Lifshitz solutions but also the $z = 1$ AdS solutions in these models break supersymmetry and are unstable. We show this explicitly in section \ref{topdown} by demonstrating that scalar modes around AdS violate the Breitenlohner-Freedman (BF) bound. The operators dual to these (unstable) scalar modes arise in the operator product expansions of the vector operators associated with the Lifshitz deformations and correspondingly are necessarily part of the consistent truncation of the bulk theory to $D$ dimensions, see section 5.2. Therefore the $z \sim 1$ Lifshitz solutions in these top down models are unstable. It would be interesting to find analogous top down solutions which are obtained from deformations of supersymmetric AdS critical points and which do not suffer from BF instabilities.

For the four-dimensional Lifshitz geometries, which are realized as solutions of the six-dimensional Romans theory, there is a second branch of solutions for which the dynamical exponent $z > 4.29$. These solutions are not connected to the unstable critical point and therefore cannot be understood in terms of marginal Lifshitz deformations. It would be interesting to understand this branch of the solutions further.

\bigskip

The plan of this paper is as follows.
In the next section we summarise the key results from \cite{Korovin:2013bua}. Assuming that the sources are position independent we extend the previous analysis to generic dimension. In section \ref{black-h} we develop the perturbation theory in $\e$ and obtain the black brane solutions in generic dimensions. In section \ref{thermo} we discuss the thermodynamics of our solutions, given an argument for and verifying the Lifshitz scaling behaviour. In section \ref{topdown} we demonstrate that the Lifshitz solutions  and the corresponding black holes \cite{Barclay:2012he} of the top-down model \cite{Gregory:2010gx} are in the same universality class as those considered in the present paper, i.e. they can be viewed as describing the ground state and a thermal state, respectively, of a relativistic CFT deformed by a vector of dimension $d$.

\section{Summary of holographic dictionary} \label{sec:HoloRen}

In this section we briefly review the holographic dictionary between bulk Lifshitz spacetimes with dynamical exponent $z = 1 + \e^2$ and the dual Lifshitz invariant field theory. We will follow the discussion in \cite{Korovin:2013bua} to which we refer the reader for more details. For use in subsequent sections, we give the renormalised action, the holographic one point functions and their Ward identities.

Note that holographic renormalization for Lifshitz solutions
was also studied previously in \cite{Ross:2009ar,Ross:2011gu,Baggio:2011cp,Griffin:2011xs,Mann:2011hg,Baggio:2011ha}. In particular, it was shown in \cite{Ross:2011gu}, using the
radial Hamiltonian formalism \cite{Papadimitriou:2004ap,Papadimitriou:2004rz},
that Lifshitz models can be holographically renormalized for any $z$.
Since these models are non-relativistic it is natural to work in the vielbein formalism \cite{Ross:2009ar}(see also \cite{Guica:2010sw}) and this is indeed what was done in \cite{Ross:2011gu}. In the current context we use instead the metric formalism \cite{deHaro:2000xn} as this is more natural when studying the theory from the perspective of the AdS critical point.

The action under consideration is
\begin{align}
\label{action}
 S_{\text{bare}} &= \frac{1}{16 \pi G_{d+1} } \int{d^{d+1} x \sqrt{-G} \Big(R + d(d-1) - \frac{1}{4} F_{\mu \nu} F^{\mu \nu}
- \frac{1}{2} M^2 A_{\mu} A^{\mu} \Big)} \nn
\\
& \qquad + \frac{1}{8 \p G_{d+1}} \int d^{d}x \sqrt{-\g}K,
\end{align}
with $M^2=d-1 + \mathcal{O}(\e^2)$, $\g$ the induced boundary metric and $K$ the trace of the second fundamental form. The associated field equations are
\begin{align}
& D_{\mu} F^{\mu \nu} = M^2 A^{\nu}, \label{Eins} \\
&R_{\mu \nu} = -d G_{\mu \nu} + \frac{M^2}{2} A_{\mu} A_{\nu} + \frac{1}{2} G^{\rho \sigma} F_{\mu \rho} F_{\nu \sigma} +
\frac{1}{4(1-d)} F^{\sigma \lambda} F_{\sigma \lambda} G_{\mu \nu}.
\end{align}
Taking the trace of the Einstein equations and plugging back into \eqref{action} the onshell action becomes
\begin{align}
\label{onshell}
S_{\text{onshell}} &= \frac{1}{16 \pi G_{d+1}} \int{d^{d+1} x \sqrt{-G} (-2d - \frac{1}{2(d-1)} F_{\mu \nu} F^{\mu \nu})} \\
& \qquad +\frac{1}{8 \pi G_{d+1}} \int{d^d x \sqrt{-\g} K}. \nn
\end{align}
It is useful to parametrize the metric and the vector field as
\begin{align}
\label{param}
&ds^2 = dr^2 + e^{2r} g_{ij} dx^i dx^j,\nn \\
&g_{ij}(x,r;\e) = g_{[0]ij}(x,r) + \epsilon^2 g_{[2]ij}(x,r) + \ldots  \\
& A_i(x,r;\e) = \epsilon e^r A_{(0)i}(x) + \ldots. \nn
\end{align}
For the metric, the notation $g_{[a]ij}$ captures the order in $\e$. When one considers the asymptotic behaviour near the conformal boundary, each of these
coefficients admits a radial expansion as well and the order in radial expansion
will be denoted (as usual) by curved parentheses. For example,
\be
g_{[0]ij}(x,r) = g_{[0](0)ij}(x) + e^{-2 r} g_{[0](2)ij}(x) + \cdots
\ee
is the asymptotic radial expansion of the metric to leading order in $\e$. Below we summarize the most general asymptotic solution given $g_{[0](0)ij}$ and
$A_{(0)i}$ as Dirichlet data.

\subsection{Asymptotic expansions}

Throughout this paper we will be interested in the case in which the background sources $g_{[0](0)}$ and $A_{(0)}$ are constant. This allows us to drop non-radial derivatives in the subsequent analysis and simplify many formulae from the \cite{Korovin:2013bua}. In particular the radial component of the vector field vanishes, $A_{r}=0$.


The results provided below hold for any dimension $d$.
The near boundary expansions of the vector field and the metric up to order $\e^2$ are as follows.
For the vector
the asymptotic expansion takes the form
\begin{align}
\label{ex}
&A_i =  e^{r} ( {\cal  A}_{(0)i} + e^{-d r} (r \tilde{\cal{A}}_{(d)i}(x)+{\cal A}_{(d)i}(x) ) +\ldots ),
\end{align}
where we will define ${\cal A}_{(0) i} = \ep A_{(0) i}$ and work perturbatively in $\ep$.
It is also useful to define ${\cal A}_{(d) i} = \ep A_{(d) i}$
and let $\tilde{{\cal A}}_{(d)}^i = \ep a_{(d)}^i$.
The logarithmic expansion coefficient is given by
\be
\label{q}
a_{(d)i} =g_{[0](d)ij} A_{(0)}^j,
\ee
where $g_{[0](a)ij}$ is defined below. 
Here and later, whenever we present asymptotic solutions,
 indices are raised to this order using the metric
$g_{[0](0)}^{ij}$, which we set to be the Minkowski metric: $g_{[0](0)ij} = \eta_{ij}$.

The coefficient $A_{(d) i}$ is left undetermined by the asymptotic analysis and
is related to the expectation value of the dual
operator. Note that the expansion coefficients depend locally on the
zeroth order expectation value of the dual stress
energy tensor $\langle T_{ij} \rangle_{[0]}$ (which is related to $g_{[0](d)ij}$ as in \eqref{zeroth}). At first sight this might appear problematic
since this coefficient is in general non-locally related to  $g_{[0](0)}$ which might lead to non-local divergences but as we review below there are in fact no non-local divergences: the counterterm action is local.

For the metric the asymptotic expansion is as follows \cite{Skenderis:1999nb,deHaro:2000xn,Korovin:2013bua}:
\begin{align}
g_{ij} = \eta_{ij}+ \e^2 r h_{[2](0)ij} + e^{-dr}\Big(\e^2 r h_{[2](d)ij} + (g_{[0](d)ij}+\e^2 g_{[2](d)ij}) \Big),
\end{align}
where the metric $g_{[0](0)ij}$ is chosen to be flat and
\be
\label{h20}
h_{[2](0)ij} = - A_{(0)i}A_{(0)j} + \frac{1}{2 (d-1)}A_{(0)k} A^k_{(0)} \eta_{ij}.
\ee
$g_{[0](d)ij}$ is traceless and divergenceless, while
\begin{align}
\label{h22}
h_{[2](d)ij} &= \frac{d}{4(d-1)} A_{(0)k} A_{(0)}^{k}g_{[0](d)ij} +\frac{1}{d} A_{(0)}^{k} g_{[0](d)kl} A_{(0)}^{l} \eta_{ij} \\&-\frac{d-1}{d}(A_{(0)i} g_{[0](d)jk} + A_{(0)j} g_{[0](d)ik})A_{(0)}^{k}, \nn \\
\tr(g_{[2](d)}) \label{trg2}
&= \frac{2}{d} A_{(0)i} A^i_{(d)}-\frac{d^2 - 2d + 2}{d^2 (d-1)}
A^i_{(0)} g_{[0](d)ij} A^j_{(0)},
\end{align}
and the divergence of $g_{[2](d)}$ vanishes.
The part of  $g_{[2](d)ij}$ which is undetermined by the asymptotic analysis is related to the expectation value of $T_{ij}$ at order $\e^2$.

\subsection{Counterterms and renormalized one-point functions}

The counterterm action, restricted to the case where the boundary metric is flat and so is $A_{(0)}$, becomes
\begin{align}
S_{\text{ct}}=S_{\text{ct}[0]}+ S_{\text{ct}[2]} &= - \frac{1}{32 \pi G_{d+1}} \int{d^d x \sqrt{-\g}} \Big(4(d-1)- \g^{ij} { A}_i { A}_j\Big). 
\end{align}
These counterterms suffice to render the action finite to order $\ep^2$, under the above restrictions.
For non-constant sources there are also other counterterms (see \cite{Korovin:2013bua}) but these do not play a role here.

The vector one-point function is
\be
\label{J}
\langle {\cal J}^i \rangle =-\frac{1}{\sqrt{-g_{[0](0)}}} \frac{\delta S_{\text{ren}}}{\delta {\cal A}_{(0)i}}=-\frac{1}{16 \pi G_{d+1}} (d {\cal A}_{(d)}^i - g_{[0](d)}^{ij} {\cal A}_{(0)j}). 
\ee
The part of the asymptotic expansion, ${\cal A}_{(d)}^i$, undetermined by asymptotics is directly related to the one-point function of the dual operator.

Now let us give the $1$-point function of the stress-energy tensor:
\be
\langle {\cal T}_{ij} \rangle = \langle T_{ij}\rangle_{[0]} + \ep^2 \langle T_{ij}\rangle_{[2]} + \cdots
\ee
with \cite{Skenderis:1999nb,deHaro:2000xn}
\be
\langle T_{ij} \rangle_{[0]} = -\frac{2}{\sqrt{-g_{[0](0)}}} \frac{\delta S_{[0]\text{ren}}}{\delta g_{[0](0)}^{ij}}
= \frac{d}{16 \pi G_{d+1} } g_{[0](d)ij} \label{zeroth}
\ee
and
\begin{align}
\label{T}
\langle T_{ij}\rangle_{[2]} 
&=\!\frac{1}{16 \pi G_{d\!+\!1}} \Big[ d g_{[2]\!(d)ij} \!-\!(A_{\!(0)i} A_{\!(d)j}\!+\!A_{(0)j} A_{(d)i})
-\! A_{\!(0)k} A^k_{\!(d)} \eta_{ij} \\&+\frac{d-1}{d}(A_{(0)i}g_{[0](d)jk} + A_{(0)j}g_{[0](d)ik})A^k_{(0)}  \nn \\
&+\frac{d^2\!-\!d\!+\!2}{2d(d-1)}A_{\!(0)}^k g_{[0](d)kl} A^l_{\!(0)}\eta_{ij}-\frac{d\!-\!2}{4(d\!-\!1)}A_{\!(0)k} A_{\!(0)}^k g_{[0](d)ij}\Big].\nn
\end{align}
Again, as expected, the expectation value of ${\cal{T}}_{ij}$ is directly related to the undetermined coefficient, $g_{(d)ij}$.

\subsection{Ward identities} \label{sec:WId2}

The holographic energy momentum tensor satisfies
\be
\label{dWard}
\nabla^j \left\langle \mathcal{T}_{ij} \right\rangle = \mathcal{A}_i \nabla_j \left\langle \mathcal{J}^{j} \right\rangle - \left\langle \mathcal{J}^{j} \right\rangle  \mathcal{F}_{ij}.
\ee
Computing the trace of the second order stress energy tensor gives the complete anomaly through order $\e^2$
\bea \label{Ward}
\langle {\cal T}^i_i \rangle
- \frac{1}{2} {\cal A}_{(0)}^i \langle {\cal T}_{ij} \rangle  {\cal A}_{(0)}^j
= {\cal A}_{(0)i} \langle {\cal J}^i \rangle.
\eea
The terms quadratic in ${\cal A}_{(0)i}$
can be thought of as a beta function contribution to the trace Ward identity \cite{Korovin:2013bua}.

\section{A Lifshitz black brane solution} \label{black-h}

In this section we construct a Lifshitz black brane solution analytically, working perturbatively in $\ep$ around the AdS neutral black brane. Working up to second order in $\ep$, the resulting black brane solution has Lifshitz asymptotics with dynamical exponent $z=1 + \ep^2$.

The leading order metric in $(d+1)$ dimensions can be expressed as
\be
ds^2 = \frac{d y^2}{c_0} - c_0 dt^2 + y^2 dx \cdot dx  ,
\label{bhmetric}
\ee
with $y \rightarrow \infty$ at the boundary and $y =  y_{h}$ at the horizon. The metric function $c_0$ is given by
\be
c_0 = y^2 (1 - {y_{h}^d}/{y^{d}}).
\ee
If a source for the vector field $\ep A_{(0) t}$ is switched on, then at order $\ep$ the bulk solution is given by the black hole metric together with a vector field $A_t (y)$ satisfying
\be
\frac{1}{d-1}\pa_{y}^2 A_t + y^{-1} \pa_y A_t - \frac{1}{c_0} A_t = 0,
\label{eomat}
\ee
subject to the condition that $A_t \rightarrow 0$ on the horizon $y = y_{h}$ and $A_{t} \rightarrow \ep A_{(0)t} y$ as $y \rightarrow \infty$.
The solution that satisfies these boundary conditions is
\bea
A_t &=& \ep A_{(0)t} a(y); \\
a(y)&=&
\frac{\p}{\sin \frac{\p}{d}} \frac{d-1}{d^2} {y}
(1-\frac{y_{h}^d}{y^d})\ {}_{2}F_{1} (\frac{1}{d}, \frac{d-1}{d}; 2; 1-\frac{y_{h}^d}{y^d}), \nn
\eea
where the normalization has been fixed for future convenience. As $y \rightarrow \infty$ this solution behaves as
\be
a(y)={y} \Big[1 - \frac{d-1}{d} \frac{y_{h}^d}{y^{d}} \log \frac{y}{y_{h}} - \frac{y_{h}^d}{y^{d}} \Big(1+\frac{d-1}{d^2} - \frac{(d-1)}{d^2}k(d)\Big) +\ldots\Big],
\ee
where we have introduced
\be
k(d)=2\g + \psi(\frac{d+1}{d}) + \psi(\frac{2d-1}{d})
\ee
to shorten formulae; $\g$ is the Euler-Mascheroni constant, and $\psi$ denotes the digamma function.

The coordinate $y$ can be changed to a domain-wall type radial coordinate as
\begin{align}
r= \int{\frac{dy}{\sqrt{c_0}}}.
\end{align}
Near the boundary $y = \infty$ this can be integrated to give the following asymptotic expansion
\begin{align}
r= \log y - \frac{1}{2 d} \frac{y_{h}^d}{y^d} + \ldots.
\end{align}
In the new coordinates the metric \eqref{bhmetric} is
\be
ds^2 = dr^2 - e^{2r} (1 + \frac{1-d}{d} y_{h}^d e^{-dr} + \ldots) d{t}^2 + e^{2r} (1 + \frac{1}{d} y_{h}^d e^{-dr} + \ldots) d{x} \cdot d{x}.
\ee
Thus at zeroth order in $\e$ we recover the well-known result
\begin{align}
\left\langle T_{tt}\right\rangle_{[0]} &= \frac{d-1}{16 \p G_{d+1}}y_{h}^d, \\
\left\langle T_{ij}\right\rangle_{[0]} &=\frac{\delta_{ij}}{16 \p G_{d+1}}y_{h}^d.
\end{align}
and the stress-energy tensor is manifestly traceless.
Also in the new variable
\begin{align}
\label{aexp}
&a(r)\!=\! e^r \Big[ 1 \!- \! \frac{d\!-\!1}{d}y_{h}^d r e^{\!-\!dr}\!-\!\frac{d\!-\!1}{d^2}\Big( \frac{2 d^2 \!+\! d \!-\!2}{2(d-1)} \!-\!k(d) \!-\! d \log y_{h} \Big)y_{h}^d e^{\!-\!dr} \Big].
\end{align}
The expansion \eqref{aexp} agrees with the earlier result from solving the field equations asymptotically; a
nontrivial check is provided by the ratio of $a_{(d)i}$ and $A_{(0)i}$ coefficients in the asymptotic expansion.
They are related to each other by \eqref{q}.
It is easy to see that this relation is in agreement with \eqref{aexp}.
Moreover, we can extract the one-point functions $\left\langle {\cal J}^t \right\rangle$ using \eqref{J} 
\begin{align}
 \left\langle {\cal J}^{t} \right\rangle &= y_{h}^d \frac{\e A_{(0)t}}{16 \p G_{d+1}}\Big(-\frac{2d-1}{2} + \frac{d-1}{d}k(d)+ (d-1) \log y_{h}\Big).
\end{align}

Next let us consider the backreaction of the vector field onto the metric at order $\ep^2$. It is convenient to parameterize the metric as follows
\be
ds^2 = \frac{d y^2}{ c(y)} - d{t}^2 c(y) b(y)^2 + y^2 d {x} \cdot d {x},
\ee
letting
\be
c (y) =c_0+ \ep^2 A_{(0) t}^2 \Delta c (y); \qquad
b(y) = 1 + \ep^2 A_{(0)}^2 \Delta b(y).
\ee
With this parametrization the Einstein equations give the change in the Ricci tensor as
\begin{align}
\ep^{-2} A_{(0)t}^{-2} \Delta R_{yy} &=
\frac{d}{c_0^2} \Delta c(y) + \frac{(2-d)}{2 (d-1)} \frac{(\pa_y a(y))^2}{c_0}; \\
\ep^{-2} A_{(0) t}^{-2}
\Delta R_{{t}{t}} &= d (\Delta c(y) \!+\! 2 c_0 \Delta b(y)) \!+\! \frac{1}{2} M^{\!2\!} a(y)^{\!2\!} \! +\! \frac{(d\!-\!2)}{2 (d\!-\!1)} c_0(\pa_y a(y))^2 ;  \\
\ep^{-2} A_{(0)t}^{-2} \Delta R_{{i} {j}} &= \delta_{ij} \frac{1}{2 (d-1)} y^2 (\pa_y a(y))^2.
\end{align}
The perturbation in the Ricci tensor computed from the metric gives
\begin{align}
\ep^{-2} A_{(0) t}^{-2} \Delta R_{yy} =& - \frac{3}{2} \pa_y \ln (c_0) \pa_y \Delta b(y)
- \pa_y^2 \Delta b(y) & \\
& - \frac{ \pa_y^2 \Delta c(y)}{2 c_0}- \frac{(d-1) \pa_y \Delta c(y)}{2 y c_0}
+ \frac{d \Delta c (y)}{c_0^2}; &\nn \\
\ep^{-2} A_{(0) t}^{-2} \Delta R_{tt} = &
d (\Delta c(y) + 2  c_0 \Delta b(y))\!+\! \frac{d\!-\!1}{2}\frac{c_0}{y} \pa_y \Delta c(y) \!+\! \frac{1}{2} c_0 \pa_y^2 \Delta c (y) &\\
 &+ y c_0 \Big( (d+2) - (\frac{d-4}{2}) \frac{y_h^d}{y^d}\Big)
\pa_y \Delta b(y)  + c_0^2 \pa_y^2 \Delta b(y) ; &\nn \\
\ep^{-2} A_{(0) t}^{-2} \Delta R_{ij} =&\delta_{ij}
\Big( \!- y \pa_y \Delta c(y) \!-\! y c_0 \pa_y \Delta b(y) - (d\!-\!2) \Delta c(y) \Big).
\end{align}
It is convenient to consider the following combination of equations:
\begin{align}
\label{EqDc}
c_0 \D R_{yy} + \frac{1}{c_0}\D R_{{t} {t}} + g^{{i} {j}} \D R_{{i}{j}},
\end{align}
which leads to a decoupled equation
\be
\label{eqDc}
\pa_y(y^{d-2}  \D c) = - \frac{1}{2} \frac{y^{d-1}}{c_0} \Big( {a^2} + \frac{c_0 (\pa_y a)^2}{(d-1)} \Big).
\ee
Using the differential equation satisfied by $a(y)$ in the form
\be
\pa_y (y^{d-1} \pa_y a) = \frac{(d-1) y^{d-1}}{c_0} a
\ee
the righthandside of the above equation can be simplified to give
\be
\pa_y (y^{d-2}  \D c)=- \frac{1}{2 (d-1)} \pa_y\left ( a \pa_y a y^{d-1} \right ).
\ee
This equation can be integrated to
\begin{align}
\D c(y) &= -\frac{(d-1)\p^2}{4 d^5 \sin^2(\frac{\p}{d})} \frac{c_0}{y^d}\ {}_{2} F_{1} (\frac{1}{d},\frac{d-1}{d};2;1 - \frac{y_{h}^d}{y^d})\times \\
&\times\Big[2d(y^d + (d-1)y_{h}^d)\ {}_{2} F_{1} (\frac{1}{d},\frac{d-1}{d};2;1 - \frac{y_{h}^d}{y^d}) \nn\\
&+ (d-1)y_{h}^d (1 - \frac{y_{h}^d}{y^d}) \ {}_{2} F_{1} (\frac{2d-1}{d},\frac{d+1}{d};3;1 - \frac{y_{h}^d}{y^d}) \Big] + c_h y_{h}^{d-2} \frac{y^2}{y^d}, \nn
\end{align}
where $c_h$ is integration constant chosen such that $\D c(y_h) = c_h$.

The asymptotic expansion for $\D c(y)$ is
\begin{align}
\label{Dc}
\D c(y) &=   y^2  \Big[ \frac{1}{2(1-d)} - \frac{(d-2)}{2 d} \frac{y_{h}^d}{y^d} \log (\frac{y}{y_{h}})\\&+\frac{y_{h}^d}{y^d} \Big(\frac{d-2}{2d^2} k(d) - \frac{d^3 - 2d^2 - 2d +2}{2d^2 (d-1)} +\frac{c_h}{y_h^2} \Big)\ldots \Big].\nn
\end{align}
Therefore,
\begin{align}
c(y) &= y^2 \Big(1-\frac{\e^2 A_{(0)t}^2}{2(d-1)} -\frac{(d-2) \e^2 A_{(0)t}^2}{2d} \frac{y_{h}^d}{y^d} \log (\frac{y}{y_{h}}) - \frac{y_{h}^d}{y^d} \\&+ \e^2 A_{(0)t}^2  \frac{y_{h}^d}{y^d} \Big(\frac{d-2}{2d^2}k(d) - \frac{d^3 - 2d^2 - 2d +2}{2d^2 (d-1)} +\frac{c_h}{y_h^2} \Big) \ldots \Big).\nn
\end{align}
In the backreacted geometry the domain-wall type radial coordinate differs from that in the original asymptotically AdS black brane spacetime. The radial coordinate is now given by
\begin{align}
 r &= \int dy \frac{1}{\sqrt{c_0 + \ep^2 A_{(0)t}^2 \D c}} \\
&= \Big(1+\frac{\e^2 A_{(0)t}^2}{4(d-1)} \Big)\log ({y}) -\e^2 A_{(0)t}^2 \frac{d-2}{ 4 d^2} \frac{y_{h}^d}{y^d} \log (\frac{y}{y_{h}}) - \frac{1}{2 d}\frac{y_{h}^d}{y^d} \nn\\&+  \frac{\e^2 A_{(0)t}^2}{4 d^3 }\Big((d-2)k(d) - \frac{2d^3 + d^2 - 10d +8}{2(d-1)} + 2d^2 \frac{c_h}{y^2_h} \Big) \frac{y_{h}^d}{y^d} + \ldots .\nn
\end{align}
The integration constant is fixed by the requirement that there is no contribution to the source at order $\e^2$, i.e. $g_{[2](0)xx} =0$.
Inverting this relation we get
\begin{align}
 &y = e^r \Big[1 + \frac{1}{2d} y_{h}^d e^{-dr} + \ldots
+\e^2 A_{(0)t}^2 \Big(-\frac{r}{4(d-1)} + \frac{(3d-4)}{8 d^2}y_{h}^d r e^{-dr} \\& + (\frac{d^2 + d-4 -(d-2)k(d)}{4 d^3} -\frac{c_h}{2d y_h^2}-\frac{d-2}{4d^2} \log y_{h})y_{h}^d e^{-dr} + \ldots \Big) \Big] \nn
\end{align}
and hence
\begin{align}
 &g_{{x}{x}} \!=\! e^{2r} \Big[1 \!+\! \frac{1}{d} y_{h}^d e^{-dr} + \!\ldots \!
+\e^2 A_{(0)t}^2 \Big(\!-\!\frac{r}{2(d\!-\!1)} \!+\! \frac{(3d\!-\!2)(d\!-\!2)}{4 d^2 (d-1)}y_{h}^d r e^{-dr} \\&+ (\frac{d^2 + d-4 -(d-2)k(d)}{2 d^3} -\frac{c_h}{d y_h^2}-\frac{d-2}{2d^2} \log y_{h})y_{h}^d e^{-dr} + \ldots \Big) \Big].\nn
\end{align}
Now we can check that
\begin{align}
h_{[2](0){x} {x}} &= \frac{1}{2(d-1)} A_{(0)t} A_{(0)t} g_{[0](0)}^{tt} g_{[0](0)xx},
\end{align}
and
\begin{align}
h_{[2](d){x}{x}} = y_{h}^d \frac{(3d\!-\!2)(d\!-\!2)}{4 d^2 (d-1)} A_{(0)t} A_{(0)t},
\end{align}
in agreement with \eqref{h20} and \eqref{h22} respectively.


Next, we solve for $\D b(y)$:
\begin{align}
c_0 \pa_y \Delta b(y) &= - \pa_y \Delta c(y) - \frac{d-2}{y} \Delta c (y) -  \frac{1}{2 (d-1)} y (\pa_y a(y))^2,
\end{align}
or
\begin{align}
\label{EqDb}
\pa_y \Delta b(y)=\frac{y a^2(y)}{2 c_0^2}.
\end{align}
Integrating this equation close to the boundary we find
\begin{align}
\D b (y) = & \Big[\frac{1}{2} \log y + \frac{d-1}{d^2} \frac{y_{h}^d}{y^d}\log (\frac{y}{y_{h}}) + \tilde{b} +\frac{d-1}{d^3} (2 - k(d)) \frac{y_{h}^d}{y^d} + \ldots \Big],
\label{Dbexp}
\end{align}
where $\tilde{b}$ is an integration constant.
This gives the time component of the metric in domain-wall coordinates
\begin{align}
g_{tt} &= - c_0 - \ep^2 A_{(0)t}^2 (\D c + 2 c_0 \D b) \\
&= -e^{2r} + \frac{d-1}{d} y_{h}^d e^{(2-d)r} + \e^2 A_{(0)t}^2 e^{2 r} \Big[(-1+\frac{1}{2(d-1)}) r \nn\\& +\frac{ (d-2) (7d-6)}{4 d^2} y_{h}^d r e^{-dr}  + \Big(-\frac{(d-1)(d-6)}{2d^2} \log y_{h}\nn\\&-\frac{(d-1)(d-6)k(d)}{2d^3}+\frac{(d-1)(d-4) (d+3)}{2d^3} -\frac{d-1}{d} \frac{c_h}{y_h^2}\Big) y_{h}^d e^{-dr}  \Big]+\ldots. \nn
\end{align}
Again the source should not be modified and therefore $\tilde{b} = 1/(4 (d-1))$.
We see that
\begin{align}
h_{[2](0)tt} &= (-1+\frac{1}{2(d-1)})A_{(0)t} A_{(0)t}, \\
h_{[2](d)tt} &= y_{h}^d \frac{ (d-2) (7d-6)}{4 d^2} A_{(0)t} A_{(0)t},
\end{align}
both in agreement with \eqref{h20} and \eqref{h22} correspondingly.

Using \eqref{T} we compute 
\begin{align}
\left\langle T_{tt} \right\rangle_{[2]} &\!=\! y_{h}^d\frac{A_{\!(0)t} A_{\!(0)t}}{16 \p G_{d\!+\!1}} \Big(\frac{2 d-1}{4}-\frac{d-1}{2} \Big(\log y_h \!+\!\frac{k(d)}{d} \!+\! \frac{2 c_h}{y_h^2}\Big)\Big) ,\\
\left\langle T_{ij} \right\rangle_{[2]} &\!=\! y_{h}^d\frac{A_{\!(0)t} A_{\!(0)t}}{16 \p G_{d\!+\!1}} \Big( -\frac{1}{4(d-1)}+\frac{ \log y_{h}}{2} \!+\! \frac{k(d)}{2d} \!-\!\frac{ c_h}{y_h^2} \Big) \delta_{ij}.
\end{align}
It is straightforward to check that the Ward identities \eqref{Ward} are satisfied.

\section{Thermodynamics} \label{thermo}

In this section we will discuss the thermodynamics of the black brane solution, working in Euclidean signature for convenience.
To define the mass we need to take into account the fact that the stress-energy tensor is not conserved by itself, but satisfies a non-trivial Ward identity \eqref{dWard}.
Consider the current
\be
Q_j =  (\left\langle \mathcal{T}_{ij} \right\rangle - \mathcal{A}_i  \left\langle \mathcal{J}_{j} \right\rangle) \xi^i,
\ee
where $\xi^i$ is such that $\nabla^{0}\xi^{0}=z$, $\nabla^{a}\xi^{b}= \delta^{ab}$, $\nabla^{a}\xi^{0}= \nabla^{0}\xi^{a}=0$.
Using the Ward identity we get
\begin{align}
\nabla^j Q_j &=  - \left\langle \mathcal{J}^{j} \right\rangle (\xi^i \nabla_i \mathcal{A}_j + \mathcal{A}_i \nabla_j \xi^i) + \left\langle \mathcal{T}_{ij} \right\rangle \nabla^j \xi^i \\
&= - z \left\langle \mathcal{J}^{t} \right\rangle \mathcal{A}_t + z \left\langle \mathcal{T}_{t}^t \right\rangle +\left\langle \mathcal{T}_{i}^i \right\rangle, \nn
\end{align}
which is precisely the Ward trace identity.

Therefore the current $Q^i$ is conserved and following \cite{Papadimitriou:2005ii} we can define the conserved mass as
\be
M = \int_{t=const} \sqrt{g} (\left\langle T_{tt}\right\rangle - {\cal A}_{t} \left\langle {\cal J}_{t}\right\rangle).
\ee
Note that expressions for the one-point functions \eqref{J} and \eqref{T} remain the same upon analytic continuation to Euclidean signature as explained in \cite{Skenderis:2009nt}.

The horizon location at order $\ep^2$ is shifted to
\be
y_{0} = y_{h}(1 -{ \frac{1}{d y_{h}^2}} \e^2 A_{(0)t}^2 c_{h}).
\ee
The Hawking temperature obtained from the requirement of no conical singularity in Euclidean signature is shifted to
\be
T = \frac{d y_{h}}{4 \pi} \left (1 + \ep^2 A_{(0)t}^2 \left [
\frac{(d-3)}{d} \frac{c_h}{y_{h}^2}  + \frac{1}{d} \frac{\pa_y \Delta c (y_h)}{y_{h}}  + b_h \right ] \right )
\ee
with $b_h = \Delta b (y=y_{h})$ and the entropy defined as the area of the horizon becomes
\be
S = \frac{ V y_{h}^{d-{1}}}{ {4} G_{d+1}} (1 - \ep^2 A_{(0)t}^2 \frac{(d-1)}{{d}} \frac{c_h}{y_{h}^2}),
\ee
with $V$ being the regulated volume of the horizon. The constant $c_h$ is directly related to the position of the horizon, which is the only independent parameter characterizing the thermodynamic properties of the black brane. The derivative of $\D c$ at the horizon is
\be
\pa_y \Delta c (y_h) = -\frac{(d-1) \p^2 y_h}{2d^2 \sin^2(\frac{\p}{d})} + (2-d)\frac{c_h}{y_h}.
\ee
We compute $b_h$ using equation \eqref{EqDb} in integrated form
\be
\Delta b(y) \Big|^{y_0}_{y_h}=\int_{y_h}^{y_0}\frac{y a^2(y)}{2 c_0^2} dy,
\ee
where $y_0$ is a near-boundary cut-off. Using the expansion \eqref{Dbexp} we find
\begin{align}
\label{Db1}
b_h &= \frac{1}{4 (d-1)}+ \lim_{y \rightarrow \infty} \Big(\frac{1}{2}\log y -\int_{y_{h}}^{y}  \frac{y' a^2(y')}{2 c_0^2(y')}dy' \Big) \\
&= \frac{1}{4 (d-1)}+\frac{1}{2}\log y_{h}+ \lim_{y \rightarrow \infty}  \int_{y_{h}}^{y}\Big(\frac{1}{2 y'}-  \frac{y' a^2(y')}{2 c_0^2(y')} \Big)dy'. \nn
\end{align}
The limit on the right hand side of the last line is finite and independent of $y_{h}$. We evaluate it numerically and get in $d=2$
\be
b_h = \frac{1}{2}\log y_{h}+0.42370309... \sim \frac{\p^2}{16} - \log 2 + \frac{1}{2}+\frac{1}{2}\log y_{h}
\ee
and in $d=3$
\be
b_h = \frac{1}{2}\log y_{h}+0.22974839.... \sim - \frac{1}{2} \log 3 + \frac{4 \pi^2}{81} + \frac{7}{24}+\frac{1}{2}\log y_{h}.
\ee

More generally, the first law of thermodynamics allow us to fix that for any $d$
\be
b_h = \frac{(d-1) \p^2}{2 d^3 \sin^2(\p/d)} + \frac{k(d)}{2d} + \frac{\log y_h}{2} - \frac{2d^2-3d+2}{4d(d-1)},
\ee
which agrees with the above expressions in $d=2$ and $d=3$.

Up to order $\ep^2$ we get the following results for the thermodynamic quantities:
\begin{align}
\label{therm}
M &=\frac{ (d\!-\!1)V y_0^{d}}{16 \p G_{d+1}} \Big[1+\e^2 A_{(0)t}^2 \Big(-\frac{2d-1}{4(d-1)}  +\frac{k(d)}{2d}  + \frac{1}{2}\log y_0\Big)\Big],  \\
S &= \frac{V y_0^{d-1}}{4 G_{d+1}}, \\
\label{therm2dT}
T &= \frac{d y_0}{ 4 \p} \Big[1 + \ep^2 A_{(0)t}^2  \Big(b_h - \frac{(d-1) \p^2}{2 d^3 \sin^2(\p/d)}  \Big) \Big].
\end{align}



As a non-trivial check one can verify that the relation
\be
\label{therm2}
M = \frac{d-1}{d+z-1}TS
\ee
is satisfied. Such a relation must hold in any Lifshitz invariant theory on general grounds: in equlibrium we have $T_{\m\n} = \text{diag}(e,-p,\ldots,-p)$, where $p$ is pressure and $e = \frac{dM}{dV}$ is energy density. The fundamental thermodynamic relation implies that
\be
e + p = T s,
\ee
where $s = \frac{dS}{dV}$ is the entropy density. Invariance under Lifshitz scaling yields
\be
z e = (d-1)p.
\ee
These together imply \eqref{therm2}.

Let us make another important observation. Recall that this black brane is asymptotically Lifshitz with $z=1+\e^2 A_{(0)t}^2/2$. From \eqref{therm2dT} we can read off the scaling relation between temperature and entropy (recall that $y_0$ is the independent parameter) \cite{Bertoldi:2009dt,Bertoldi:2011zr,Barclay:2012he}
\be
 T \sim S^{\frac{z}{d-1}}.
\ee
This relation together with \eqref{therm} implies the first law of thermodynamics
\be
dM = T dS.
\ee

\subsection{On-shell action}

In this section we evaluate the Euclidean on-shell action. (Note that under the Wick rotation $t \rightarrow - i \t$ all the terms in the action change their sign. ) 

The on-shell action for the Lifshitz black brane can be expressed as a sum of terms:
\be
S_{\text{on-shell}} = S_{\text{bulk}} + S_{\text{GH}} + S_{\text{ct}}.
\ee
We begin by computing $S_{\text{bulk}}$:
\be
S_{\text{bulk}} = -\frac{1}{16 \p G} \int d^d x \int_{y_{0}}^{y_c} dy \sqrt{G} (-2d - \frac{1}{2 (d-1)} F_{\m\n} F^{\m\n})
\ee
where $y_c$ is a radial cutoff and $y_0$ is the position of the horizon. Noting that $\sqrt{G} = y^{d-1} b(y)$ we compute
\begin{align}
 S_{\text{bulk}}
&\!=\! -\frac{1}{16 \p G} \int d^d x \int_{y_{0}}^{y_c} \!dy  y^{d-1}\Big(\!-\!2d +
\ep^2 A_{(0)t}^2 \Big(\frac{(\pa_y a)^2}{(d-1)} -\!2d \D b(y)\Big)\Big) \\
&= -\frac{1}{16 \p G} \int d^d x \Big[ -2y^d (1 + \ep^2 A_{(0)t}^2 \D b(y))\Big|_{y_{0}}^{y_c}
+ \ep^2 A_{(0)t}^2 \frac{y^{d-1} a \pa_y a}{d-1}\Big|_{y_0}^{y_c} \nn\\&+\e^2 A_{(0)t}^2 y_0^d \Big(\frac{1}{2 (d-1)} + \log y_c\Big)
- 2 \ep^2 A_{(0)t}^2 y_0^d b_h \Big]. \nn
\end{align}
Here we have integrated by parts, used the field equation \eqref{EqDb} for $\D b(y)$ along with the defining equation
\eqref{eomat} for $a(y)$.

Now we move on to evaluate the Gibbons-Hawking term and the counterterms. Working in the $y$ coordinate,  the induced metric at the regulated surface satisfies
\be
\sqrt{\g} = (y_c^d - \frac{y_h^d}{2}) \Big(1 +  \ep^2 A_{(0)t}^2 \D b(y_c) + \frac{1+(y_h/ y_c)^{d}}{2 y_c^2} \ep^2 A_{(0)t}^2 \D c(y_c)\Big).
\ee
The Gibbons-Hawking term can be combined with the leading order counterterm to give
\be
 K\!-\!(d\!-\!1) = 1 \!+\! \ep^2 A^2_{(0)t} \Big(\frac{\tr(h_{[2](0)})}{2} \!-\! \frac{y_c^{-d}}{2} \tr(h_{[2](d)} \!-\!d g_{[2](d)} \!- \!g_{[0](d)}h_{[2](0)})\Big).
\ee
Thus we obtain
\begin{align}
&-\frac{1}{8 \p G_3}\int d^2 x \sqrt{\g} (K-d+1)  \\&= -\frac{1}{8 \p G_{d+1}} \int d^d x
y_c^d \Big[1-\frac{y_0^d}{2y_c^d}+ \ep^2 A_{(0)t}^2\Big( \D b(y_c) +\frac{d-3}{4(d-1)}  \nn \\&\!-\! \frac{d\!-\!1}{2d}\Big(\frac{y_0}{y_c}\Big)^d \log y_c\!+\! \Big(\frac{y_0}{y_c}\Big)^d \Big(\frac{5d\!-\!6}{4d}(\frac{k(d)}{d} + \log y_0) \!-\!\frac{7d^3 \!-\! 4 d^2 \!-\! 16 d \!+\! 12}{8 d^2(d-1)} \Big) \Big)\Big]. \nn
\end{align}
Now we compute the on-shell action term by term by plugging in asymptotic expansions for $a$, $\D c$ and $\D b$:
\begin{align}
 &S_{\text{bulk}} = -\frac{1}{8 \p G_{d+1}} \int d^d x
\Big[-y_c^d + y_0^d  - \e^2 A^2_{(0)t} y_c^d \Big(\D b(y_c)+\frac{1}{2(1-d)} \\
&\!+\frac{1\!-\!d}{d} \Big(\frac{y_0}{y_c}\Big)^d \log y_c \!+\!(d\!-\!2) \Big(\frac{y_0}{y_c}\Big)^d \Big( \frac{k(d)}{2 d^2} \!-\! \frac{(2d^2 +d-2)}{4d^2(d-1)} +\frac{\log y_0}{2d} \Big) \Big)  \Big]. \nn\\
\end{align}
The remaining contributing counterterm is
\begin{align}
\!-\frac{1}{32 \p G_{d\!+\!1}}\int d^d x \sqrt{\g} A_i A^i  &\!=\!\frac{1}{32 \p G_{d\!+\!1}}\int d^d x \e^2 A_{(0)t}^2 \Big[y_c^d \!-\!\frac{2(d\!-\!1)}{d}y_0^d \log y_c \\&-2y_0^d \Big( \frac{3d^2 + 4d -4}{4 d^2}-\frac{d-1}{d}(\frac{k(d)}{d}+\log y_0) \Big)\Big]. \nn
\end{align}
Putting all these terms together gives the free energy
\begin{align}
S_{\text{on-shell}} =\b F= -\b \frac{V y_0^d}{16 \p G_{d+1}} \Big[1 +\e^2 A_{(0)t}^2\Big(\frac{k(d)}{2d} +\frac{\log y_0}{2} - \frac{1}{4(d-1)}\Big) \Big],
\end{align}
where $\b = 1/T$.
It is a simple check that $F = M - T S$.

\section{Relation to top down solutions} \label{topdown}

In \cite{Gregory:2010gx,Braviner:2011kz,Barclay:2012he} Lifshitz solutions of Romans gauged supergravity theories were constructed and then uplifted to ten dimensional supergravities\footnote{Note however that the uplifts from six dimensions to massive IIA given in \cite{Gregory:2010gx} have typos in the Bianchi identities.}. General dynamical exponents with $z \ge 1$ were obtained. Here we will consider the limit of these solutions as $z \rightarrow 1$ and interpret them from the perspective of the dual conformal field theory of the AdS $z=1$ solution. Recently uplifts of the six-dimensional Romans theory to type IIB were found
\cite{Jeong:2013jfc} and thus these solutions may also be viewed as solutions of type IIB.

Here we will discuss mostly the Lifshitz solutions in four bulk dimensions (henceforth denoted ${\rm Li}_4$) which are obtained as solutions of the Romans gauged supergravity in six dimensions since the four-dimensional case is phenomenologically more interesting and moreover corresponding finite temperature solutions were constructed in \cite{Barclay:2012he}. An analogous discussion holds for the Lifshitz solutions in three bulk dimensions found in \cite{Gregory:2010gx} and we will summarise the properties of these solutions at the end of this section.

We begin by reviewing the equations of motion for the six-dimensional Romans theory \cite{Romans:1985tw}.  The bosonic field content
of 6D Romans' supergravity consists of the metric, $g_{AB}$,
a dilaton, $\phi$, an anti-symmetric two-form
field, $B_{AB}$, and a set of gauge vectors,
$(A_A^{(i)},{\mathcal A}_A)$ for the gauge
group $SU(2)\times U(1)$.  The bosonic part of the
action for this theory is 
\begin{eqnarray}
S &=&\int d^6 x  \sqrt{-g_6} \Bigg[
 \frac14 R_6 {- } \frac12 (\partial \phi )^2
-\frac{e^{-\sqrt{2} \phi}}{4}\,
\left( {\cal H}^2 +F^{(i)2}\right)
{ - } \frac{e^{2\sqrt{2}\phi}}{12} G^2  \\
&&{ - } \frac18 \, \varepsilon^{ABCDEF}\,
B_{AB} \left( {\cal F}_{CD} {\cal F}_{EF}
+{m} B_{CD} {\cal F}_{EF}
+\frac{{ m}^2}{3} B_{CD} B_{EF}
+F^{(i)}_{CD} F^{(i)}_{EF} \right)\nonumber\\
&&+\frac{1}{8}\left( g^2 e^{\sqrt{2} \phi}
+4 g { m} e^{-\sqrt{2} \phi}-{ m}^2
 e^{-3 \sqrt{2} \phi}
\right) \Bigg ] , \nonumber 
\end{eqnarray}
where $g$ is the gauge coupling, $m$ is the mass of the two-form field $B_{AB}$,
${\cal F}_{AB}$ is a U(1) gauge field strength, $F^{(i)}_{AB}$ is a nonabelian
SU(2) gauge field strength, $G_{ABC}$ is the field strength of the two-form and ${\cal H}_{AB}
= {\cal F}_{AB} + { m} B_{AB}$.
Spacetime indices $A,B,...$ run from $0$
to $5$, and $\varepsilon$ is the Levi-Civita tensor density. 

Varying the action gives the Einstein equation
\be
\begin{aligned}
R_{AB} &= 2 \partial_A \phi \partial_B \phi
- \frac12 g_{AB} V(\phi) {+}
e^{2\sqrt{2}\phi} \left(G_A^{\,\,\,CD}
G_{BCD} - \frac16 g_{AB}
G^2 \right) \\
& + e^{-\sqrt{2}\phi} \left(2 \H_A^{\,\,C} \H_{BC}
+ 2 F_A^{iC} F_{BC}^i
- \frac14 g_{AB} \left(\H^2
+ (F^i)^2 \right) \right),
\end{aligned}
\label{6deom1}
\ee
and the following matter equations of motion
\begin{align} \label{6deom5}
& \Box \phi = - \frac12 \frac{\partial V}{\partial\phi} {+ }  \frac13
\sqrt{\frac{1}{2}} e^{2\sqrt{2}\phi} G^2 - \frac12
\sqrt{\frac12} e^{-{\sqrt{2}\phi}} \left(\H^2
+ (F^{(i)})^2\right) \\
&\nabla_B \left( e^{-\sqrt{2}\phi} \H^{BA} \right)
= {} \frac16 \, \epsilon^{ABCDEF} \H_{BC} G_{DEF} \nn \\
&\nabla_B \left( e^{-\sqrt{2}\phi} F^{(i)BA} \right)
= { } \frac16 \, \epsilon^{ABCDEF} F^{(i)}_{BC} G_{DEF} \nn \\
&\nabla_C \left(e^{2\sqrt{2}\phi} G^{CAB} \right)
\!=\! {m} e^{-\sqrt{2}\phi} \H^{AB} {+} \frac14 \,
\epsilon^{ABCDEF} \left(\H_{CD} \H_{EF}
\!+\! F_{CD}^{(i)} F_{EF}^{(i)} \right) \,, \nn
\end{align}
where we have defined the scalar potential function as
\be
V(\phi) = \frac{1}{4} \left ( g^2 e^{\sqrt{2}\phi}
+ 4 {m} g e^{-\sqrt{2}\phi}
-{m}^2 e^{-3\sqrt{2}\phi} \right ) \,.
\ee
The equations of motion admit a solution which is ${\rm Li}_4 \times H_2$, with $H_2$ a hyperboloid
\be
ds^2 = L^2 \left ( - y^{2z} dt^2 + y^2 dx \cdot dx + \frac{dy^2}{y^2} \right ) + a^2 ds^2(H_2),
\ee
where $L$ is the curvature radius of ${\rm Li_4}$ and $a$ is the curvature radius of the hyperboloid, with $ds^2(H_2)$ denoting the unit radius metric. Relative to \cite{Barclay:2012he} the signature has been changed to mostly plus, to fit the conventions of this paper, and the radial coordinate is denoted $y$ in accordance with the earlier sections. In the Lifshitz solutions, the scalar field is constant, $\phi = \phi_0$, and the field configurations are
\bea
F^{(3)}_{ty} &=& q B L^3 e^{\sqrt{2} \phi_0} y^{z-1}; \qquad F^{(3)}_{H_2} = q \eta_{H_2},  \label{fel}  \\
B_{x_1 x_2} &=& \frac{B}{2} L^3 y^2. \nn
\eea
Here $\eta_{H_2}$ is the volume form of the hyperboloid. The Lifshitz solutions exist only if the parameters are related by algebraic equations which are expressed in terms of the following quantities
\be
\begin{aligned}
&\hat B = L B e^{\sqrt{2}\phi_0} \qquad Q = L e^{-\phi_0/\sqrt{2}} q/a^2\\
& \hat g = L g e^{\phi_0/\sqrt{2}}
\qquad\hat a = a/L \qquad
\hat{m}= L\, {m}\, e^{-3\phi_0/\sqrt{2}} \,.
\end{aligned}
\label{eq:LifshitzRescalings}
\ee
and hence one gets
\be
\begin{aligned}
&\hat B^2 = z-1 \qquad \hat g^2= 2 z(4+z) &
& \frac{\hat{m}^2}{2}=\frac{6+z \mp 2\sqrt{2(z+4)}}{z}
\\
&Q^2=\frac{ (2+z) (z-3)\pm 2 \sqrt{2(z+4)}}{2 z} &
& \frac{1}{\hat a^2} = 6+3z \mp 2\sqrt{2(z+4)} \, .
\end{aligned}
\label{eq:lifsolns}
\ee
From here onwards we will set the curvature radius $L$ to be one and the integration constant $\phi_0$ to be zero, in which case the hatted quantities are the same as those without hats. Note however that $Q = q/a^2$. 
Flux quantization may impose restrictions on the allowed values of $z$, forcing $z$ to take discrete values but in what follows we will not discuss these restrictions.

\subsection{Lifshitz with $z \sim 1$}

There are two branches of Lifshitz solutions, but only the upper sign solutions are connected to the AdS solution with $z=1$ and it is this branch that we will focus on here. When $z=1$, $B = 0$ and
\bea
{ m} &=& \sqrt{10}  - 2; \qquad Q^2 = \sqrt{10} - 3; \\
g^2 &=& 10; \qquad \qquad \frac{1}{a^2} = 9 - 2 \sqrt{10}. \nn
\eea
At this critical point
\be
V(0) = 9 - \sqrt{10}.
\ee
Now letting
\be
B^2 \equiv \e^2 = z - 1,
\ee
with $\e$ small, we note that at order $\e$ the solution is the leading order $AdS_4 \times H_2$ together with an $F^{(3)}_{ty}$ flux of order $\e$ and a $B_{x_1 x_2}$ flux also of order $\e$. At order $\e^2$ both parts of the six dimensional metric are changed (note that the radius of the hyperboloid is corrected) and the flux along the hyperboloid is also corrected at this order.

To interpret this limit, it is useful to look at the spectrum around the $AdS_4 \times H_2$ background. We will not need the complete spectrum in what follows; it suffices to look at the following decoupled modes.  Switch on perturbations around the background
\be
A^{(3)}_{\mu} = a_{\mu} (x^{\rho}); \qquad
B_{\mu \nu} = b_{\mu \nu} (x^{\rho}),
\ee
where $x^{\mu}$ denote the $AdS_4$ coordinates. Such perturbations do not depend on the $H_2$ coordinates and are therefore singlet modes from the perspective of the Kaluza-Klein reduction over this (compact) space. Linearising the equations of motion around the background these modes decouple from all other linear perturbations but are coupled to each other via the equations
\bea
\bar{\nabla}_{\mu} f^{\mu \nu} &=&  \frac{1}{3} Q \ep^{\nu \rho \sigma \tau} g_{\rho \sigma \tau}; \\
\bar{\nabla}_{\rho} g^{\rho \mu \nu} &=& m^2 b^{\mu \nu}  + Q \ep^{\mu \nu \sigma \tau} f_{\sigma \tau}. \nn
\eea
Here $\bar{\nabla}$ denotes the $AdS_4$ covariant derivative, $\ep^{\mu \nu \rho \sigma}$ is the covariant  epsilon on $AdS_4$ and $(f_{\mu \nu}, g_{\mu \nu \rho})$ are the curvatures of the vector and tensor field perturbations, respectively. These equations should be supplemented by the divergence constraint on the tensor field, $\bar{\nabla_{\mu}} b^{\mu \nu} = 0$. This system of equations has the degrees of freedom
of a massive vector field: define
\be
c_{\mu} = \frac{1}{3!} \ep_{\mu \nu \rho \sigma} g^{\nu \rho \sigma}.
\ee
Closure of the three form $g$ implies that $c_{\mu}$ is divergenceless. Denoting the curvature of $c$ as $f_c^{\mu \nu}$ the coupled equations of motion reduce to
\be
\bar{\nabla}_{\mu} f_{c}^{\mu \nu} = 2 c^{\nu};
\ee
with this massive vector field strength being related to the gauge field strength and the tensor field as
\be
f_{c}^{\mu \nu} = 2 Q f^{\mu \nu} - \frac{1}{2}  m^2 \ep^{\mu \nu \rho \sigma} b_{\rho \sigma}.
\ee
Comparing with \eqref{Eins}, we note that such an equation describes a massive vector field with $M^2 = 2 =  d-1$ and thus this mode is precisely the vector mode considered in earlier sections. Working to order $\e$, the Lifshitz solution is therefore indeed a deformation of the dual conformal field by the time component of a massive vector operator of dimension three: expanding \eqref{fel} to first order in $\e$ one can extract
\be
c_t = - \e y. \label{1flux}
\ee
(Note that with our conventions $\ep_{ty12} = -{y^2}$.)

\subsection{Backreaction of massive vector field}

Working perturbatively around $z=1$, the backreaction of the massive vector preserves Lifshitz invariance with $z = 1 + \ep^2$ when evaluated on the AdS background. If one adds a massive vector perturbation to an asymptotically AdS background, such as a black brane, the resulting solution will only be asymptotically Lifshitz. Moreover, the backreaction of the massive vector will be non-trivial not just on the four-dimensional metric, but also on the two scalar fields, which will now run.

To analyse the backreaction of the massive vector field at order $\ep^2$ it is useful to first reduce the six-dimensional equations to a set of four-dimensional equations using the following ansatz for the fields \cite{Braviner:2011kz}.  The six-dimensional Einstein frame metric is expressed as
\be
ds^2 = e^{\frac{1}{2} \chi} g_{\mu \nu} dx^{\mu} dx^{\nu} + e^{- \frac{1}{2} \chi} a^2 ds^2 (H_2),
\ee
where the factors are chosen such that $g_{\mu \nu}$ is an Einstein frame metric in four dimensions. For the other six-dimensional fields,
\bea
\phi &=& \phi(x); \qquad B_{AB} = b_{\mu \nu} (x) + b(x) \eta_{(H^2) ab}; \\
F^{(3)}_{AB} &=& f_{\mu \nu}(x) + \gamma \eta_{(H^2) ab}. \nn
\eea
Here $b(x)$ is a scalar field but $\gamma$ is a constant and ${\cal F} = 0$. The reduced action is then \cite{Braviner:2011kz}:
\begin{align}
S &= a^2 \int d^4 x \sqrt{-g} \Big[\frac{1}{4}R  -\frac{e^{-\sqrt{2}\phi -\c/2}}{4}(f_{\m\n} f^{\m\n} + m^2 b_{\m\n} b^{\m\n} ) \\
& - \frac{e^{2 \sqrt{2} \phi -\c}}{12} g_{\m\n\r} g^{\m\n\r}\!-\!\frac{1}{16} (\nabla \c)^2 \!-\! \frac{1}{2} (\nabla \phi)^2 \!-\!\frac{e^{2 \sqrt{2} \phi +\c}}{2 a^4} (\nabla b)^2 \!{-}\frac{e^{\c}}{2 a^2} \!+\! \frac{e^{\c/2}}{2} V(\phi) \nn \\
& - \frac{1}{2a^4}e^{-\sqrt{2}\phi +3\c/2} (m^2 b^2 + \g^2) \!-\!\frac{\e^{\m\n\r\s}}{8 a^2} ({ 2} m^2 b b_{\m\n} b_{\r\s} \!+\!{2} b f_{\m\n} f_{\r\s} +{4}\g b_{\m\n} f_{\r\s})\Big]. \nn
\end{align}
The interactions in the action above imply that the backreaction of the massive vector will be non-trivial not just on the four-dimensional metric, but also on the two scalar fields, which will run.
At order $\ep^2$ Lifshitz invariance is preserved but at this order other operators, as well as the stress energy tensor, are affected. As discussed in \cite{Korovin:2013bua}, the extra fields in the consistent truncation, beyond the metric and massive vector, relate to additional terms occurring in the OPE between the vector operator and the stress energy tensor. One could analyse a more general system of this type using the techniques of \cite{Korovin:2013bua} and this paper.

However, for the specific system under consideration, analysing the expansion in $\ep$ in detail is less interesting for the following reason.
Linearizing the equations of motion for the scalars $\phi$ and $\chi$ around the AdS solution one obtains
\bea
\Box \phi &=&  - \frac{1}{4} (g^2 + 4 m g - 9 m^2) \phi + Q^2 \phi - \frac{3 Q^2}{2 \sqrt{2}} \c \\
&=& (16 - 6 \sqrt{10} ) \phi - \frac{3(\sqrt{10} -3)}{2\sqrt{2}}\c; \nn \\
\Box \chi &=& (\frac{4}{a^2} - V(0) + 9 Q^2) \chi - (2 V'(0) + 6 \sqrt{2} Q^2)  \phi \\
&=&2\sqrt{10} \c - 4 \sqrt{2} ({\sqrt{10}-3}) \phi. \nn
\eea
Diagonalizing this system we find that the masses of the two independent scalar modes are
\be
m_1^2 = -2.99... ; \qquad
m_2^2 = 6.34...
 \ee
and thus the eigenmodes do not satisfy the Breitenlohner-Freedman bound $m^2\geq -9/4$; this was also observed in \cite{Braviner:2011kz,Barclay:2012he}.
Therefore these scalars correspond to instabilities of the system: the original AdS critical point is not supersymmetric and it is not stable. 

Turning now to the Lifshitz solution, we note that these unstable scalars run in the finite temperature solution.
Although these unstable modes prevent us from giving clear dual interpretation of this particular system, we have shown that it belongs to the same universality class of models discussed in \cite{Korovin:2013bua}. It would be interesting to find $z \sim 1$ Lifshitz solutions in string theory which are obtained from deformations of supersymmetric CFTs and which do not suffer from such instabilities. Note that the second branch of Lifshitz solutions found in \cite{Gregory:2010gx} have dynamical exponents $z > 1$ and are not connected to the unstable $z=1$ critical point; these have been argued to be the stable branch  \cite{Braviner:2011kz,Barclay:2012he}.

\subsection{Three-dimensional Lifshitz geometries}

In this section we briefly summarize the interpretation of the ${\rm Li}_3 \times H_2$ solutions of Romans ${\cal N} = 4$ gauged supergravity in five dimensions found in \cite{Gregory:2010gx}.

The bosonic field content of the Romans theory \cite{Romans:1985ps} consists of the metric, $g_{AB}$,
a dilaton, $\phi$, two antisymmetric tensors $B^{\alpha}_{AB}$, and a set of gauge vectors,
$(A_A^{(i)},{\mathcal A}_A)$ for the gauge
group $SU(2)\times U(1)$.  The bosonic part of the
action for this theory is 
\begin{eqnarray}
S &=&\int d^5 x  \sqrt{-g_5} \Bigg[
 \frac14 R_5 {- } \frac12 (\partial \phi )^2
-\frac{\xi^2}{4}\,
\left( {B^{\alpha}}^2 +F^{(i)2}\right)
{ - } \frac{\xi^{-4}}{4} {\cal F}^2  \\
&&{ - } \frac14 \, \varepsilon^{ABCDE}\,
 \left( \frac{1}{g_1} \e_{\alpha\beta} B^{\a}_{AB} \nabla_{C} B^{\beta}_{DE} -F^{(i)}_{AB} F^{(i)}_{CD} {\cal A}_E \right)\nonumber\\
&&+\frac{1}{8} g_2 \left( g_2 \xi^{-2} + 2 \sqrt{2} g_1 \xi 
\right) \Bigg ] , \nonumber 
\end{eqnarray}
where we have defined $\xi = e^{\sqrt{2} \phi/\sqrt{3}}$. Here $g_1$ and $g_2$ are the gauge couplings
for $U(1)$ and $SU(2)$ respectively.
${\cal F}_{AB}$ is a U(1) gauge field strength and $F^{(i)}_{AB}$ is a nonabelian
SU(2) gauge field strength. Spacetime indices $A,B,...$ run from $0$
to $4$, and $\varepsilon$ is the Levi-Civita tensor density. 

Just as in the six-dimensional case there exist Lifshitz solutions
\be
ds^2 = L^2 (-y^{2z} dt^2 + y^2 dx^2 + \frac{dy^2}{y^2}) + a^2 ds^2 (H_2),
\ee
where $L$ is the curvature radius of ${\rm Li_3}$ and $a$ is the curvature radius of the hyperboloid, with $ds^2(H_2)$ denoting the unit radius metric.  In the Lifshitz solutions, the scalar field is constant and can be set to zero; for notational simplicity we will also set the curvature radius $L$ to one in the metric above. There are two distinct classes of Lifshitz solutions. The first has the following fluxes
\be
{\cal F}_{ty} = - \alpha_1  y^{z-1}; \qquad F^{(3)}_{H_2} = a^2  \gamma \eta_{H_2},   \qquad
F^{(3)}_{yx} = \beta_2.
\ee
Here $\eta_{H_2}$ is the volume form of the hyperboloid. This Lifshitz solution requires\footnote{Note that there are typos in equations (3.32) and (3.39) of \cite{Gregory:2010gx}: $\hat{g}_1^2$ should read $\hat{g}_1 \hat{g}_2$.}
\bea
&& \alpha_1^2 = \frac{1}{2} z (z-1); \qquad \beta_2^2 = \frac{1}{2} (z-1); \qquad \gamma^2 = \frac{z}{4}; \\
&& g_2^2 = -2 z^2 + 3z + 2; \qquad g_1 g_2 = \frac{1}{\sqrt{2}} (2z^2 +z + 1); \qquad a^2 = \frac{2}{3z}. \nn
\eea
In the second class of Lifshitz solutions the fluxes are
\be
{F}^{(3)}_{ty} = - \alpha_2  y^{z-1}; \qquad F^{(3)}_{H_2} = a^2  \gamma \eta_{H_2},   \qquad
{\cal F}_{yx} = \beta_1.
\ee
Here $\eta_{H_2}$ is the volume form of the hyperboloid. This Lifshitz solution requires
\bea
&& \alpha_2^2 = \frac{1}{2} z (z-1); \qquad \beta_1^2 = \frac{1}{2} (z-1); \qquad \gamma^2 = \frac{z}{4}; \\
&& g_2^2 = 2 z^2 + 3z - 2; \qquad g_1 g_2 =\sqrt{2}(1+ z); \qquad a^2 = \frac{2}{3z}. \nn
\eea
Both solutions reduce in the $z=1$ limit to the same AdS critical point. Reality of the gauge couplings requires that $1 \le z \le 2$.

Linearizing around the AdS solution the following fluctuations form a decoupled system:
\be
\delta {\cal F}_{\mu \nu} = f_{\mu \nu}(x^{\rho}); \qquad
\delta F^{(3)}_{\mu \nu} =  f^{(3)}_{\mu \nu}(x^{\rho}) ,
\ee
where $x^{\mu}$ denote AdS coordinates, with the linearised equations of motion being
\be
\bar{\nabla}_{\nu} f^{\nu \mu} = -\e^{\mu \rho \sigma} f^{(3)}_{\rho \sigma};  \qquad
\bar{\nabla}_{\nu} f^{(3) \nu \mu} = - \e^{\mu \rho \sigma} f_{\rho \sigma},
\ee
with $\bar{\nabla}_{\mu}$ the $AdS_3$ covariant derivative and $\e^{\mu\nu \rho}$ the three-dimensional covariant Levi-Civita. As previously, we can define
\be
c_{\mu} = \frac{1}{2} \ep_{\mu \nu \rho} f^{\nu \rho}
\ee
such that $c_{\mu}$ is divergenceless and is a massive vector
\be
\bar{\nabla}_{\mu} f_{(c)}^{\mu \nu} = c^{\nu},
\ee
where $f_{(c) \mu \nu} \equiv - 2 f^{(3)}_{\mu \nu}$ is the curvature of $c_{\mu}$.
This is precisely the mass given above \eqref{Eins}. Looking now at the Lifshitz solutions, we see that when $z = 1 + \e^2$ the two classes of solutions reduce to the following perturbations about the $AdS_3$ background, respectively:
\be
c_{t} = \frac{1}{\sqrt{2}} \ep y;
\qquad c_{x} = \frac{1} {\sqrt{2}} \ep y.
\ee
Therefore, the first class corresponds to a deformation by the time component of the massive vector while the second class corresponds to a deformation by the spatial component of the massive vector. Note that only in three dimensions a deformation by the special component of the massive vector is consistent with Lifshitz symmetry -- in higher dimensions such deformation breaks the rotational symmetry. 

As in the six-dimensional models, the backreaction at order $\e^2$ generically induces other fields in addition the metric and massive vector system, since the consistent truncation to three dimensions involves additional scalar fields. However, working out this backreaction in detail is not necessary because one can already show that the system has BF instabilities. Perturbing around AdS, the dilaton together with the following breathing mode of the metric
\be
ds^2 = e^{2 \chi (x^{\mu})} \left ( - y^2 dt^2 + y^2 dx^2 + \frac{dy^2}{y^2} \right ) + e^{- \chi(x^{\mu})} ds^2 (H_2)
\ee
form a decoupled system.
The diagonalized masses of these scalar modes are
\be
m_{\pm}^2 = \frac{3}{2} \pm \frac{\sqrt{33}}{2},
\ee
with the mode such that $m_{-}^2 < - 1$ violating the BF bound. Therefore just as in the six-dimensional models the AdS critical point is unstable.

\section{Conclusions} \label{sec:con}
In this paper we have constructed analytically asymptotically Lifshitz black branes with dynamical exponents$z=1 +\e^2$. Using the holographic dictionary developed in \cite{Korovin:2013bua} and extended here to arbitrary dimension we obtained the thermodynamic properties of these neutral black branes analytically. In particular, we argued for and verified the scaling relation between temperature and entropy: $T \sim S^{\frac{z}{d-1}}$. In the non-relativistic theory care must be taken to identify the correct conserved quantity needed to define the mass $M$.

We showed that the $z \sim 1$ solutions in the top-down models of \cite{Gregory:2010gx,Barclay:2012he} belong to the same universality class as those analysed in \cite{Korovin:2013bua} and in the present paper, i.e. they can be viewed as deformations of relativistic fixed points by a dimension $d$ vector  operator $J_t$. Unfortunately, these models suffer from  Breitenlohner-Freedman type instabilities, which makes the study of these systems challenging. It would be interesting to look for other string theory embeddings of $z \sim 1$ Lifshitz geometries which do not have such instabilities.

\section*{Acknowledgments}

This work is part of the research program of the Stichting voor Fundamenteel Onderzoek der Materie (FOM), which is financially supported by the Nederlandse Organisatie voor Wetenschappelijk Onderzoek (NWO). YK and KS acknowledge support via an NWO Vici grant. KS and MT acknowledge support from a grant of the John Templeton Foundation. The opinions expressed in this publication are those of the authors and do not necessarily reflect the views of the John Templeton Foundation.

\addtocontents{toc}{\protect\vspace*{\baselineskip}}





\addtocontents{toc}{\protect\vspace*{\baselineskip}}


\providecommand{\href}[2]{#2}\begingroup\raggedright\endgroup


\end{document}